\documentclass{article}
\usepackage{spconf,amsmath,graphicx}
\usepackage{booktabs}
\usepackage[table]{xcolor}
\usepackage{comment}
\usepackage{amsfonts}
\usepackage{caption}
\usepackage{url}
\usepackage{hyperref}

\title{FusDom: Combining In-Domain and Out-of-Domain Knowledge for Continuous Self-Supervised Learning}
\name{Ashish Seth$^{1*}$ \quad Sreyan Ghosh$^{2*}$ \quad S. Umesh$^1$ \quad Dinesh Manocha$^2$ \thanks{$^*$ These authors contributed equally to this work.}}
\address{$^1$IIT Madras, India, $^2$University of Maryland, College Park, USA}
\begin{document}
%
\maketitle
%

\begin{abstract}
Continued pre-training (CP) offers multiple advantages, like target domain adaptation and the potential to exploit the continuous stream of unlabeled data available online. However, continued pre-training on out-of-domain distributions often leads to catastrophic forgetting of previously acquired knowledge, leading to sub-optimal ASR performance. This paper presents FusDom, a simple and novel methodology for SSL-based continued pre-training. FusDom learns speech representations that are robust and adaptive yet not forgetful of concepts seen in the past. Instead of solving the SSL pre-text task on the output representations of a single model, FusDom leverages two identical pre-trained SSL models, a teacher and a student, with a modified pre-training head to solve the CP SSL pre-text task. This head employs a cross-attention mechanism between the representations of both models while only the student receives gradient updates and the teacher does not. Finally, the student is fine-tuned for ASR. In practice, FusDom outperforms all our baselines across settings significantly, with WER improvements in the range of 0.2 WER - 7.3 WER in the target domain, while retaining the performance in the earlier domain\footnote{\url{https://github.com/cs20s030/fusdom}}.
\end{abstract}




%
\section{Introduction}
\label{sec:intro}

In the recent past, Self-Supervised Learning (SSL) has shown impressive performance on a variety of vision \cite{chen2020simple}, text \cite{devlin2018bert}, speech \cite{baevski2020wav2vec,hsu2021hubert,liu2020mockingjay}, and audio tasks \cite{ghosh2022decorrelating}. The primary goal is to learn representations from unlabeled data to learn high-level features that can transfer well across various tasks. In the past couple of years, the Spoken Language Processing (SLP) community has developed several sophisticated algorithms that achieve state-of-the-art (SOTA) performance on popular benchmarks \cite{yang2021superb}. A primary real-world application of SSL is to overcome the data scarcity problem for under-represented languages \cite{zhang2023google}.


Continued SSL pre-training proves to be an effective solution in many real-world use cases, including domain adaptation to the low-resource target domain \cite{hsu2021robust,chen2020recall,aghajanyan2020better} and exploiting the continuous stream of unlabeled data online to keep the model's knowledge up-to-date. However, continued SSL pre-training leads to catastrophic forgetting of past knowledge \cite{purushwalkam2022challenges} due to data that violates the IID assumption of optimization algorithms \cite{bousquet2003introduction}. Forgetting past knowledge learned from large-scale pre-training leads to sub-optimal ASR performance in both the current and previous domains.
\vspace{1mm}

\begin{figure}[t]
    \centering
    \includegraphics[width=0.45\textwidth]{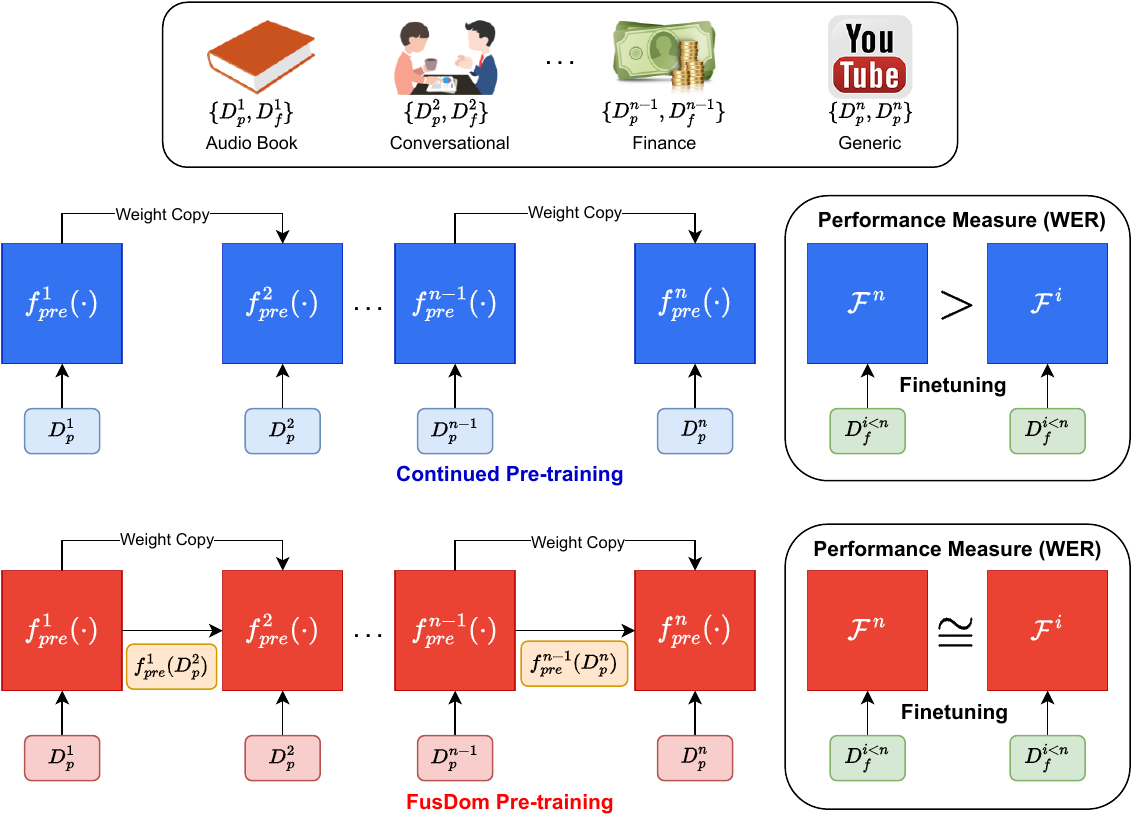}
    \caption{\small Illustration of \textbf{FusDom}. FusDom facilitates continuous SSL on multiple new distinct domains without forgetting knowledge about past domains. As a result, the resultant model achieves optimal ASR performance in the current and all previous domains.}
    \label{fig:exp_pic}
\end{figure}

\begin{figure*}[t]
    \centering
    \includegraphics[width=1.0\textwidth]{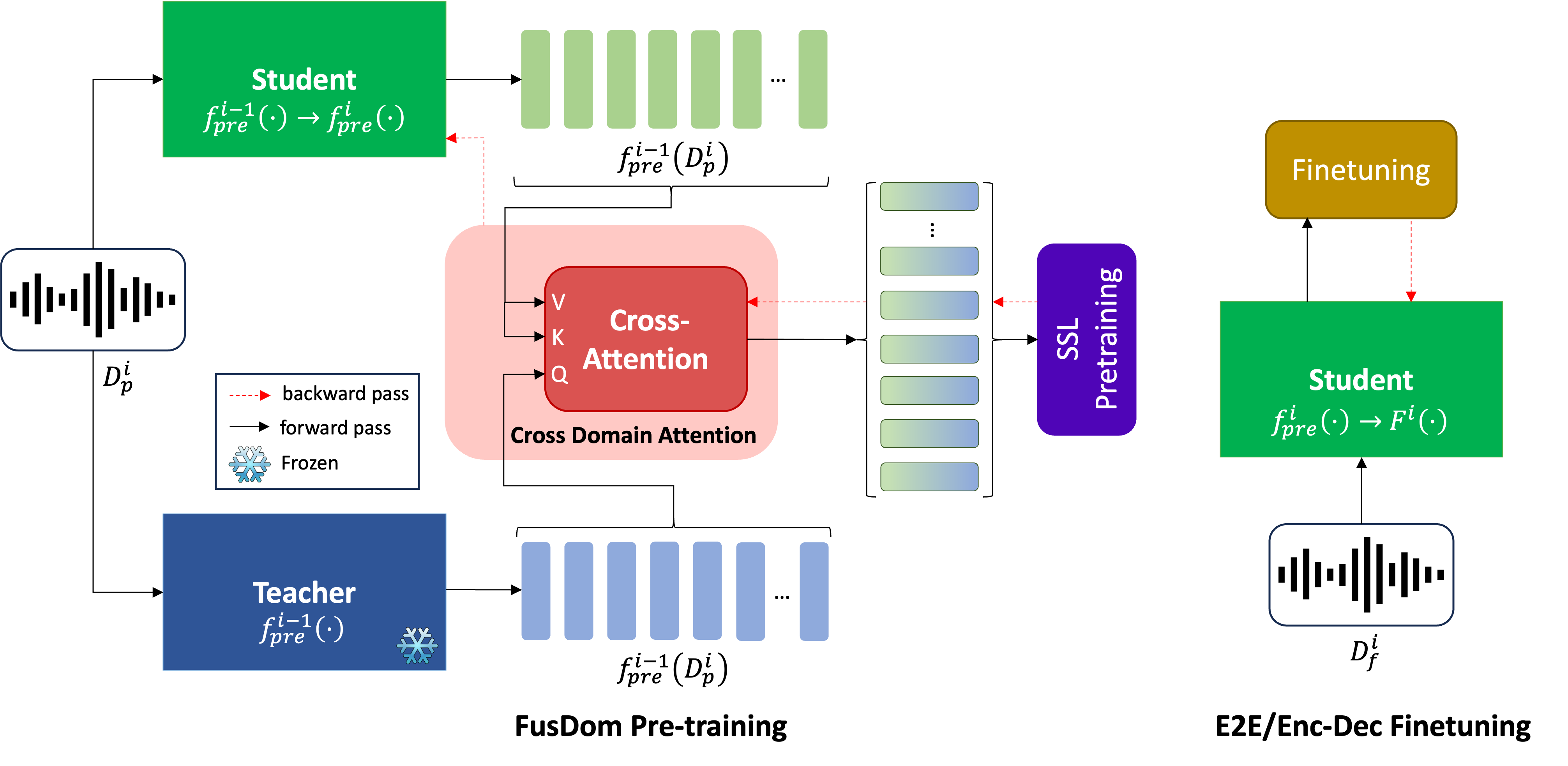}
    \caption{\small Illustration of \textbf{FusDom}. FusDom employs two identical models, a student and a teacher, for SSL-based CP (instead of 1 in \textit{vanilla} CP). For CP with target dataset $\mathcal{D}_{p}^{i}$, we initialize our student and teacher with $f^{i-1}_{pre}(.)$, which comes from the previous stage of pre-training, pre-trained on $\mathcal{D}_{p}^{i-1}$. The modified pre-training head $h_{pre}$ consists of a standard transformer block with cross-attention between representations of the student and the teachers. Precisely, in  $h_{pre}$, the teacher representations act as the query (Q) and the student representations act as the Key (K) and Value (V). Only the student receives gradient updates for SSL pre-training, while the teacher does not. Finally, the student model is either fine-tuned end-to-end using CTC or used as a feature extractor for fine-tuning an Encoder-Decoder model.}
    \label{fig:fusdom_pic}
\end{figure*}

{\noindent \textbf{Main Contributions.}} In this paper, we propose FusDom, a simple and novel continued pre-training (CP) strategy for pre-training existing SSL models on non-IID data. FusDom brings the best of both worlds by simultaneously adapting a pre-trained model to the target downstream domain and avoiding forgetting past knowledge it has learned with large-scale SSL pre-training. To achieve this, FusDom employs two identical copies of an SSL model and leverages a novel SSL pre-training head to solve the pre-text task for CP. Of these two models, only one receives gradient updates (the student), while the other is always frozen (the teacher). The novel pre-training head employs a transformer layer with cross-attention, where the queries come from the teacher, and the keys and values come from the student. Effectively, the in-domain model representations solve the pre-text in an out-of-domain-aware fashion that eventually helps retain past acquired knowledge. To build FusDom, we are inspired by the normal human learning process, where humans leverage past acquired knowledge to infer and learn new and unseen concepts. We build FusDom on the core heuristic that neural networks have enough capacity to store information about every domain it sees in continued SSL \cite{lai2021parp}. We perform an extensive empirical evaluation to prove the efficacy of FusDom in various settings and show that FusDom achieves relative Word Error Rate (WER) gains in the range of 1.2-7.2 over our baselines on target domain ASR while retaining the performance in the previous domains.








\section{Related Work}
{\noindent \textbf{SSL in Speech.}} Throughout the past decade, researchers have proposed several SSL algorithms that achieve new SOTA performance on several SLP benchmarks \cite{panayotov2015librispeech,yang2021superb}. Some of the most common ones are based on contrastive learning \cite{baevski2020wav2vec}, clustering \cite{hsu2021hubert}, or reconstruction \cite{liu2020mockingjay}. Despite its success in pushing benchmark performance, SSL models suffer from several fundamental problems which, to the best of our knowledge, lack sufficient research in the speech community. For example, SSL models suffer from catastrophic forgetting when fine-tuning and pre-training domains differ \cite{chen2020recall}. Continued pre-training on the downstream dataset has shown to be a promising direction to avoid forgetting by adapting to the target domain \cite{gururangan-etal-2020-dont}. However, effectively performing continued pre-training is difficult due to problems like over-fitting the training data and catastrophic forgetting \cite{purushwalkam2022challenges}. We acknowledge that continued pre-training overall is a relatively understudied problem in literature and specifically in speech representation learning, where the benefits and drawbacks of it are under-explored.
\vspace{1mm}

{\noindent \textbf{Continual Learning.}} One popular area of research that handles catastrophic forgetting of neural networks in the transfer learning paradigm is continual learning \cite{de2021continual}. Compared to text and vision, continual learning for speech representation learning is a relatively under-explored area, leaving much to explore \cite{yang2022online,sadhu2020continual}. Additionally, most prior work focuses on supervised-only settings, ignoring continual learning of models learned with SSL. Different from existing work, FusDom takes a first step toward solving catastrophic forgetting in the continued SSL pre-training paradigm.

\section{Methodology}

{\noindent \textbf{Problem Formulation.}} Fig. \ref{fig:fusdom_pic} shows a clear pictorial representation of our proposed approach. . Let's say we have an upstream model $f_{pre}(.)$ and a linear stream of $n$ unlabeled datasets $\mathcal{D}_p \in \{\mathcal{D}_{p}^{i}, \cdots, \mathcal{D}_{p}^{n}\}$ where each $\mathcal{D}_{p}^{i}$ = ($\mathbf{X}^{i}_{pre}$) comes from a different domain than the previous. These unlabeled datasets are employed to perform SSL on $f_{pre}(.)$ sequentially, in any order, where at each step, the resultant model can be denoted as $f^{i}_{pre}(.)$. Additionally, each unlabeled dataset in $\mathcal{D}_{p}^{i}$, also has a corresponding downstream ASR dataset $\mathcal{D}^{i}_f$=($\mathbf{X}^{i}_{targ}$,$\mathbf{Y}^{i}_{targ}$). Our primary aim is to obtain a final upstream model $f^{n}_{pre}(.)$, which, when fine-tuned on any downstream ASR dataset $\mathcal{D}^{i}_f$, achieves optimal ASR performance, irrespective of the order in which the unlabeled datasets was shown to $f_{pre}(.)$. We denote this fine-tuned model, fine-tuned on $\mathcal{D}^{i}_f$ as $\mathcal{F}^{i}$. We either fine-tune the final pre-trained model $f^{n}_{pre}(.)$ using CTC or use it as a feature extractor to fine-tune a conformer-based Encoder-Decoder model. FusDom proposes an effective methodology for continuous SSL to prevent the model from forgetting previous domains seen in the earlier stage.

\subsection{Continued Pre-training with FusDom}
\label{sec:fusdom}

During continued pre-training, FusDom tries to avoid forgetting past knowledge by learning target-domain representations that are aware of past knowledge. To achieve this, FusDom employs a standard transformer block with cross-attention between multiple representations as the pre-training head, which we denote as $h_{pre}$. $h_{pre}$ receives its input from two similar copies of pre-trained SSL model $f_{pre}^{i}(.)$. We call one of these models a student and the other a teacher. Precisely, the queries for cross-attention in $h_{pre}$ come from the teacher, and the keys and values come from the student. Only the student receives gradient updates during SSL pre-training, while the teacher does not. For simplicity, let's denote the student representations as $\mathbf{S}$ $\in$ $\mathbb{R}^{d \times M}$ and the teacher representation as $\mathbf{T}$ $\in$ $\mathbb{R}^{d \times M}$. Formally, we can denote Cross-Domain Attention or $\mathbf{CDA}$ as:

\begin{equation}
    \mathbf{CDA}(\mathbf{S}, \mathbf{T})=\operatorname{softmax}\left(\frac{\left[\mathbf{W}_{\mathbf{q}_{\mathbf{i}}} \mathbf{T}\right]^{\top}\left[\mathbf{W}_{\mathbf{k}_{\mathbf{i}}} \mathbf{S}\right]}{\sqrt{d / M}}\right)\left[\mathbf{W}_{\mathbf{v}_{\mathbf{i}}} \mathbf{S}\right]^{\top}
\end{equation}

where \{$\mathbf{W}_{\mathbf{k}_{\mathbf{i}}}$, $\mathbf{W}_{\mathbf{q}_{\mathbf{i}}}$, $\mathbf{W}_{\mathbf{v}_{\mathbf{i}}}$\} $\in$ $\mathbb{R}^{d/M \times h}$ denote the query, key, and value weight matrices, respectively, for the $i^{th}$ attention head. Finally, the output of the Cross Domain Attention layer is passed through the standard feed-forward with a skip connection and a non-linear activation. Finally the output representation of the pre-training head $h_{pre}$ is now $\mathbf{F}$ $=$ ($\mathbf{f_0}$,$\mathbf{f_1}$, $\cdots$, $\mathbf{f_{m-1}}$). For simplicity, we make the model solve the same SSL pre-text task that it solved during the earlier pre-training stage for all our experiments.
\vspace{1mm}

 \subsection{Downstream Fine-tuning for ASR}
\label{sec:asr}

Next, we fine-tune $f_{pre}^{i}(.)$ on any target dataset, preferably from one of the domains already seen during the various stages of pre-training. For fine-tuning, we either employ End-to-End CTC Fine-tuning where we adjust all weights of $f_{pre}^{i}(.)$ by introducing a linear CTC head and subsequently optimizing the model with the CTC loss or use $f_{pre}^{i}(.)$ as a frozen feature extractor for fine-tuning a Conformer-based Encoder-Decoder. For the latter,  we jointly optimize CTC and attention-based auto-regressive losses, as \cite{watanabe2017hybrid}. 




\section{Experimental Setup}
\label{sec:experimental}


{\noindent \textbf{Datasets.}} Details on individual datasets used for all our experiments can also be found in Table \ref{tab:my_label_dataset}. \textbf{(1)} \textbf{MSR.} The MSR speech corpus~\cite{inproceedings} consists of about $\approx$150 hours of labeled ASR data for three Indian languages, namely, Gujarati, Tamil, and Telegu. The utterances are sourced from \textit{human conversations}. \textbf{(2)} The Gramvani ASR dataset~\cite{bhanushali2022gram} consists of 100 hours of labeled ASR data, with a 100 / 5 / 3 hour train-dev-test split. The utterances are sourced from \textit{telephonic conversations} in Hindi of varying regional dialects. \textbf{(3)} The SwitchBoard dataset~\cite{godfrey1992switchboard} consists of 330 hours of labeled ASR data sourced from \textit{telephonic conversations} in English. For our experiments, we sample a 30-hour training split. \textbf{(4)} The Wall Street Journal dataset~\cite{paul1992design} consists of 80 hours of labeled ASR data sourced from \textit{read speech} of financial news in the Wall Street Journal.
\vspace{1mm}

\begin{table}[t]
\caption{Detailed Statistics of datasets used in our experiments. \textbf{Type} refers to Conversational or Read speech.}
    \centering
    \resizebox{1.0\columnwidth}{!}{\begin{tabular}{lcccc}
    \toprule
    \textbf{Dataset}&\textbf{Language}&\textbf{Domain}&\textbf{Type}&\textbf{Duration}\\
    & & & &(train, dev, test)\\
    \toprule
    MSR & Gujarati & General& Conv.& 40hr, 5hr, 5hr\\
    MSR & Tamil & General& Conv.&40hr, 5hr, 5hr\\
    MSR & Telugu & General& Conv.&40hr, 5hr, 5hr\\
    Gramvani (GV) & Hindi & Call Cent.& Conv.&100hr, 5hr, 3hr\\
    SwitchBoard (SWBD) & English & Call Cent.& Conv. &30hr, 5hr, N.A.\\
    Wall Street Journal (WSJ) & English & Finance & Read & 80hr, 1.1hr, 0.4hr\\\hline
    
    \toprule
    
    \end{tabular}}
    
    \label{tab:my_label_dataset}
\end{table}

    
    
    

\begin{table*}[t]
\small
\centering
  \caption{Comparison of FusDom ASR results with our baselines on both Enc-Dec and E2E evaluation settings. All results are in the format of \textbf{dev / test}. $\mathcal{R}$ and $\mathcal{C}$ indicate Read and Conversational Speech. Domain Map refers to the source pre-training $\rightarrow$ CP domain.}
  \label{tab:results}
 \resizebox{1.0\textwidth}{!}{
  \begin{tabular}{c c c c c c c c c}
    \toprule
    \textbf{Pretrained} & \textbf{Downstream} & \textbf{Domain Map} &  \multicolumn{2}{c}{\textbf{No Cont. Pretrain}} & \multicolumn{2}{c}{\textbf{Vanilla Cont. Pretrain}} & \multicolumn{2}{c}{\textbf{FusDom}}\\
    \textbf{Model}& \textbf{Dataset} & (\emph{Source} $\rightarrow$ \emph{Target}) & \cellcolor{green!20} Enc-Dec & \cellcolor{blue!15} E2E & \cellcolor{green!20} Enc-Dec & \cellcolor{blue!15} E2E & \cellcolor{green!20} Enc-Dec & \cellcolor{blue!15} E2E\\
    \toprule
    XLSR-300 & GV\textsubscript{Hindi} & $\text{General}_\mathcal{R} \rightarrow \text{Call Cent.}_\mathcal{C}$ & \cellcolor{green!20} 32.7 / 32.5 & \cellcolor{blue!15} 37.3 / 37.0 & \cellcolor{green!20} 31.6 / 31.4 & \cellcolor{blue!15} 35.3 / 35.0 & \cellcolor{green!20}\textbf{29.9 / 28.7} & \cellcolor{blue!15}\textbf{32.2 / 32.0}\\
    XLSR-300 & MSR\textsubscript{Gujarati} & $\text{General}_\mathcal{R} \rightarrow \text{General}_\mathcal{C}$ & \cellcolor{green!20} 21.7 / 28.5 & \cellcolor{blue!15} 24.4 / 32.3 & \cellcolor{green!20} 21.3 / 27.2 & \cellcolor{blue!15} 22.1 / 30.3 & \cellcolor{green!20} \textbf{21.2 / 26.6} & \cellcolor{blue!15} \textbf{21.4 / 29.4}\\
    XLSR-300 & MSR\textsubscript{Tamil} & $\text{General}_\mathcal{R} \rightarrow \text{General}_\mathcal{C}$ & \cellcolor{green!20} 28.1 / 27.7 & \cellcolor{blue!15} 33.4 / 32.1 & \cellcolor{green!20} 27.8 / 26.9 & \cellcolor{blue!15} 32.2 / 31.2 & \cellcolor{green!20} \textbf{26.8 / 26.7} & \cellcolor{blue!15} \textbf{29.3 / 29.2}\\
    XLSR-300 & MSR\textsubscript{Telugu} & $\text{General}_\mathcal{R} \rightarrow \text{General}_\mathcal{C}$ & \cellcolor{green!20} 28.3 / 28.8 & \cellcolor{blue!15} 34.1 / 32.8 & \cellcolor{green!20} 28.0 / 28.3 & \cellcolor{blue!15} 32.6 / 32.0 & \cellcolor{green!20} \textbf{27.6 / 27.2} &\cellcolor{blue!15} \textbf{29.1 / 28.3}\\
    Vakyansh & GV\textsubscript{Hindi} & $\text{General}_\mathcal{R} \rightarrow \text{Call Cent.}_\mathcal{C}$ & \cellcolor{green!20} 34.5 / 34.3 & \cellcolor{blue!15} 33.2 / 34.2 & \cellcolor{green!20} 32.7 / 32.5 & \cellcolor{blue!15} 31.7 / 31.5 & \cellcolor{green!20} \textbf{31.6 / 31.4} & \cellcolor{blue!15} \textbf{31.0 / 31.1}\\
    Wav2Vec2-Lb & SWBD\textsubscript{English} & $\text{General}_\mathcal{R} \rightarrow  \text{Call Cent.}_\mathcal{C}$ & \cellcolor{green!20} 39.1 / N.A & \cellcolor{blue!15} 22.2 / N.A. & \cellcolor{green!20} 36.2 / N.A. & \cellcolor{blue!15} 20.4 / N.A. & \cellcolor{green!20} \textbf{32.4 / N.A.} & \cellcolor{blue!15} \textbf{15.2 / N.A.}\\
    Wav2Vec2-Lb & WSJ\textsubscript{English} & $\text{General}_\mathcal{R} \rightarrow \text{Finance}_\mathcal{R}$ & \cellcolor{green!20} 12.4 / 11.6 & \cellcolor{blue!15} 11.4 / 10.9 & \cellcolor{green!20} 11.3 / 10.6 & \cellcolor{blue!15} 10.5 / 10.0 & \cellcolor{green!20} \textbf{11.1 / 10.2} &\cellcolor{blue!15} \textbf{9.6 / 9.2}\\
    \bottomrule    
        
  \end{tabular}
  }
\end{table*}

\begin{table*}[t]
\small
\centering
  \caption{Comparison of FusDom ASR when CP on diverse target domains (SWBD, WSJ) and finetuned on source domains (Libri 10hr) with our baselines on both Enc-Dec and E2E evaluation settings. All results are in the format of \textbf{dev-clean / test-clean}}
  \label{tab:results_source}
 \resizebox{1.0\textwidth}{!}{
  \begin{tabular}{c c c c c c c c c}
    \toprule
    \textbf{Pretrained} & \textbf{Downstream} & \textbf{Cont. Pretrained} &  \multicolumn{2}{c}{\textbf{No Cont. Pretrain}} & \multicolumn{2}{c}{\textbf{Vanilla Cont. Pretrain}} & \multicolumn{2}{c}{\textbf{FusDom}}\\
    \textbf{Model}& \textbf{Dataset} & \textbf{Dataset} & \cellcolor{green!20} Enc-Dec & \cellcolor{blue!15} E2E & \cellcolor{green!20} Enc-Dec & \cellcolor{blue!15} E2E & \cellcolor{green!20} Enc-Dec & \cellcolor{blue!15} E2E\\
    \toprule
    XLSR-300 & Libri\textsubscript{English} & SWBD\textsubscript{English} & \cellcolor{green!20} \textbf{13.7 / 15.2} & \cellcolor{blue!15} \textbf{10.3 / 10.3} & \cellcolor{green!20} 15.6 / 17.2 & \cellcolor{blue!15} 12.6 / 12.5 & \cellcolor{green!20} 13.8 / 15.3 & \cellcolor{blue!15}10.4 / 10.4\\
    XLSR-300 & Libri\textsubscript{English} & WSJ\textsubscript{English} & \cellcolor{green!20} \textbf{13.7 / 15.2} & \cellcolor{blue!15}  \textbf{10.3 / 10.3} & \cellcolor{green!20} 14.2 / 16.4  & \cellcolor{blue!15} 12.1 / 16.6 & \cellcolor{green!20} 13.9 / 15.4 & \cellcolor{blue!15}10.5 / 10.4\\
    Wav2Vec2-Lb  & Libri\textsubscript{English} & SWBD\textsubscript{English} & \cellcolor{green!20} 15.8 / 20.7 & \cellcolor{blue!15} 12.7 / 17.8 & \cellcolor{green!20} 20.8 / 25.3 & \cellcolor{blue!15} 18.3 / 23.6 & \cellcolor{green!20}\textbf{15.2 / 20.4} & \cellcolor{blue!15}\textbf{12.3 / 17.4}\\
    Wav2Vec2-Lb  & Libri\textsubscript{English} & WSJ\textsubscript{English} & \cellcolor{green!20} 15.8 / 20.7 & \cellcolor{blue!15} 12.7 / 17.8 & \cellcolor{green!20} 18.8 / 22.8 & \cellcolor{blue!15} 15.2 / 20.1 & \cellcolor{green!20}\textbf{15.6 / 20.4} & \cellcolor{blue!15}\textbf{12.6 / 17.7}\\

    \bottomrule    
        
  \end{tabular}
  }
\end{table*}

\begin{table}[t]
\small
\centering

  \caption{Comparing the downstream performance of FusDom vs. Vanilla CP for sequential CP on diverse domains in E2E evaluation. All results are in format \textbf{Vanilla CP / FusDom} test set WER}
  \label{tab:results_seq}
   \resizebox{0.48\textwidth}{!}{

  \begin{tabular}{c c c c c c}
    \toprule
    \textbf{Pretrained} & \textbf{Cont. Pretrain} &  \multicolumn{3}{c}{\textbf{Downstream}}\\
    \textbf{Model}& \textbf{Order} & \textbf{Libri} & \textbf{SWBD} &  \textbf{WSJ} \\
    \toprule
    XLSR-300 & SWBD $\rightarrow$ WSJ & \cellcolor{blue!15} 15.8 / \textbf{11.0} & \cellcolor{blue!15} 12.8 / \textbf{10.7} & \cellcolor{blue!15} 9.3 / \textbf{8.1} \\
    XLSR-300 & WSJ $\rightarrow$ SWBD & \cellcolor{blue!15} 13.4 / \textbf{10.8} & \cellcolor{blue!15} 11.7 / \textbf{9.8} & \cellcolor{blue!15} 9.7 / \textbf{9.2} \\
    Wav2Vec2-Lb & SWBD $\rightarrow$ WSJ & \cellcolor{blue!15} 21.0 / \textbf{18.3} & \cellcolor{blue!15} 16.1 / \textbf{13.6} & \cellcolor{blue!15} 9.8 / \textbf{9.1} \\
    Wav2Vec2-Lb & WSJ $\rightarrow$ SWBD & \cellcolor{blue!15} 24.1 / \textbf{18.7} & \cellcolor{blue!15} 15.0 / \textbf{12.7} & \cellcolor{blue!15} 10.1 / \textbf{9.7} \\
\bottomrule
        
  \end{tabular}}
  
\end{table}
{\noindent \textbf{SSL Pre-trained models.}} For our experiments, we employ either of these three pre-trained models: \textbf{(1)} \textbf{Wav2Vec2-Libri-960:} We use the base variant of Wav2Vec2 \cite{baevski2020wav2vec} pre-trained on 960 hours of LibriSpeech \cite{panayotov2015librispeech}. The model has $\approx$95M learnable parameters. \textbf{(2)} \textbf{XLSR-300:}~\cite{conneau2020unsupervised} We use the variant of XLSR with $\approx$300M learnable parameters. This model is pre-trained on VoxPopuli, MLS, CommonVoice, BABEL, and VoxLingua107 for learning cross-lingual speech representation. \textbf{(3)} \textbf{Vakyansh.}~\cite{gupta2021clsril} The Vakyansh model for pre-trained using the contrastive learning objective, similar to \cite{baevski2020wav2vec}, on 4200 hours of Hindi Data from the read-speech domain. The model is built on the base variant of wav2vec-2.0, which has $\approx$95M learnable parameters.
\vspace{1mm}

{\noindent \textbf{Baselines.}} We compare FusDom with: \textbf{(1)} \textbf{No Continued Pre-training.} This baseline follows the most common ASR fine-tuning pipeline wherein we use the pre-trained SSL model without any continued pre-training. \textbf{(2)} \textbf{Vanilla Continued Pre-training.} This baseline employs an additional step over generic SSL pre-training by performing continued pre-training on the target domain.

{\noindent \textbf{Hyperparameters.}} For Continued Pre-training, we train our wav2vec-2.0 base SSL pre-trained model for 100 epochs. For a fair comparison, with our continued pre-training baseline, we also pre-train the FusDom-based model for a total of 100 epochs. We train FusDom with a learning rate of $5e^{-4}$ using Adam optimizer. Our conformer-based encoder-decoder model has 12 encoder layers and 6 decoder layers. We train our models with a learning rate of $1.5e^{-3}$, batch size of 64, and for a total of 100 epochs.

\section{Results and Analysis}
Table \ref{tab:results} presents a performance comparison of FusDom with our baseline methods on dataset splits mentioned in Table \ref{tab:my_label_dataset}. Our experiments mimic real-world scenarios, where the target domain for CP has fewer resources and differs from the source domain, as shown in the Domain Map column of Table \ref{tab:results}. On average, FusDom surpasses the baselines by reducing WER by 0.2-7.3 in the Encoder-Decoder (Enc-Dec) setup and 1.1-7.0 in the End-to-End (E2E) setup. Specifically, in the Enc-Dec setup, relative WER improvements of 6.1\% and 6.0\% are achieved for the dev and test sets when compared to vanilla CP, and improvements of 10.2\% and 10.1\% when compared to no CP. In the E2E setup, relative WER improvements of 12.2\% and 11.9\% are observed on the dev and test sets compared to vanilla CP, and improvements of 16.9\% and 16.7\% when compared to no CP.

Table \ref{tab:results_source} highlights a performance comparison of FusDom with our baseline methods when CP on diverse domains and finetuned on the source domain. For CP we use \emph{SWBD}, \emph{WSJ}, and for finetuning XLSR-300 and Wav2Vec2-Lb., we use \emph{Libri} 10hr split as our source domain. On average, FusDom gives similar performances as compared to no CP for E2E and Enc-Dec finetuning setups with a slight increase of 0.1-0.6 in absolute WER when compared to vanilla CP with an increase of 0.5-5 in absolute WER. This shows that FusDom is more efficient than vanilla CP in retaining previous domain knowledge. We also show the effect of sequential CP as shown in Fig \ref{fig:exp_pic} in Table \ref{tab:results_seq}, where FusDom outperforms vanilla CP with a decrease of 0.7-4.8 absolute WER on the E2E evaluation.



\section{Conclusion and Future Work}
This paper proposes FuseDom, a novel methodology to continue pre-training an SSL model on a stream of unlabelled non-IID data. FuseDom avoids catastrophic forgetting by learning to solve the pre-text task with representations that are prior knowledge aware. In practice, FuseDom improves downstream ASR performance over all our baselines by a significant margin. As part of future work, we would like to build better learning systems for more effective continued pre-training and perform a layer-wise analysis of the information learned by FusDom to quantify forgetting.

\bibliographystyle{IEEEbib}
\bibliography{strings,refs}

\end{document}